\documentstyle[12pt]{article}
\advance\hoffset by -1.1truecm
\advance\voffset by -1.0truecm
\begin{document}
\title{Path Integral Approach to Noncommutative Spacetimes}
\author{Gianpiero Mangano\thanks{E-mail: mangano@na.infn.it} \\ \\
INFN, Sezione di Napoli, and
Dipartimento di Scienze Fisiche, \\
Universit\`a di Napoli {\it Federico II}, Italy.\\}
\date{\empty}
\maketitle
\begin{abstract}
We propose a path integral formulation of noncommutative 
generalizations of
spacetime manifold in even dimensions, characterized by a length
scale $\lambda_P$. The commutative case is obtained in the limit
$\lambda_P=0$.\\
\\
PACS numbers: 04.60.Gw, 04.60.-m, 02.40-k.
\end{abstract}


\newpage
\section{Introduction}

It is commonly believed that the picture of spacetime as a smooth 
pseudo--Riemannian 
manifold should 
breakdown at most at the Planck scale $\lambda_P\sim 10^{-33}~ cm$, 
due to the 
effects of quantum gravity. 
Many arguments on operational limits on position and time 
measurements have been considered in
literature \cite{acv} - \cite{mm}, suggesting that our 
description of spacetime as a collection of points
equipped with suitable topological and metric structures, should be 
modified in a way which
is reminiscent of the quantization of phase space in ordinary quantum 
mechanics. In this case
too, in fact, the classical phase space, seen as the joint spectrum 
of commuting position and
momentum operators, turns into its fuzzy quantum version, where the 
notion of points is 
replaced, in a somehow intuitive way, by fundamental cells of area 
$\hbar$. This quantization
procedure can be formalized by adopting a {\it dual} 
point of view, namely
by considering the algebra of smooth functions over the phase
space, 
generated by positions and momenta. This algebra contains,
using Gelf'and--Naimark reconstruction theorem, all informations on 
the underlying space. Switching on the
Planck constant then amounts to consider the noncommutative algebra 
still generated by positions
and momenta, considered now as non commuting elements and with usual 
commutation relations. 

Quantum phase space is perhaps the most famous 
example of what is now usually called Noncommutative Geometry. While
at the classical level the phase space can be described either {\it
geometrically}, i.e. defining its structure as a topological manifold, or
algebraically, its noncommutative quantum version can only be defined via the
algebra of observables generated by positions and momenta, since the notion 
of ordinary manifold is lost.
The algebraic description of noncommutative geometries has recently 
received new insights in particular due to Connes \cite{connes}.

In this paper we will assume, as first considered long ago \cite{sny},
and as suggested 
by the previous considerations, that the 
effect of gravity at very short distances is such to render positions 
and time $X^\mu$ 
non commutative
\begin{equation}
[X^\mu,X^\nu]= i \lambda_P^2 Q^{\mu \nu}
\label{1}
\end{equation}
where the antisymmetric tensor $Q^{\mu \nu}$, depending in general 
on $X^\mu$, should
be eventually given by a satisfactory theory of quantum gravity. 
The spacetime is described
by the $*$-algebra ${\cal A}$ generated by regular representations
of Eq.(\ref{1}), 
whereas the notion of
points is now lost. Here and in the following we will use upper 
case for the elements
of ${\cal A}$ and lower one for their commutative limit, so for 
example, $F(X)$ is an
element of ${\cal A}$ and $f(x)$ is instead an ordinary 
smooth function over the commutative spacetime manifold. 
Many ansatzs for Eq.(\ref{1})
have been considered, in different contexts, \cite{dfr}, \cite{mm},
\cite{madore}, \cite{km}. 
They predict uncertainty relations
among position and time operators which in turn imply that quantum field 
theories built upon these
generalized spacetimes have a more regular behaviour at 
small distances, and, in particular, 
may be free of
ultraviolet divergences. See for example Refs. \cite{dfr}, 
\cite{km} - \cite{gkp}.  

The key observation of the present paper is that the analogy with 
ordinary quantum mechanics 
and quantization of phase space can also be pursued to establish a path 
integral formulation of
these noncommutative geometries, taking the point of view that a 
class of linear functionals
over ${\cal A}$, which turn into evaluation maps in the commutative 
limit,
can be expressed as integrals, with a suitable measure, of ordinary 
functions over the classical, commutative spacetime 
manifold. It is well known in fact that
quantization of phase space
can be introduced either in the canonical approach, 
namely by defining 
a representation for positions and momenta as 
self-adjoint operators on a 
separable Hilbert space, or via
Feynman sum over paths in classical phase space. In this case, to 
evaluate matrix elements of
time ordered polynomials in position and momenta one should average 
over all trajectories with
fixed boundary conditions, $q(t_0)=q_0$ and $q(t_f)=q_f$,
with measure $\exp(i S /\hbar)$. For one degree of freedom
\begin{equation}
<q_f,t_f| T[Q(t_1) P(t_2)... ]|q_0, t_0>=\int D[q] D[p]~ e^{i 
S/\hbar} 
~(q(t_1) p(t_2)... )
\label{3} 
\end{equation}
where $T$ denotes time ordering and the action $S$ is the integral 
over 
the considered path of the one-form $p dq - H(p,q) dt$. 
Equation (\ref{3}) defines a set of linear functionals over the algebra 
generated by
$Q$ and $P$ in the Heisenberg representation.
In particular for $q_0=q_f$,  
of the two terms in $S$, the first one measures for each closed
trajectory $\gamma$ in the phase space the area of the surface 
$\Gamma$ 
with boundary $\gamma$ normalized to $\hbar$, i.e. the
surface integral of the 
symplectic form $dp \wedge dq$ over $\Gamma$. 
Actually, the picture of quantum phase space as composed by
fundamental cells of area $\hbar$ is already
emerging from this term,
independently of which particular dynamical system we are 
considering, defined by a particular
choice for the Hamilton function $H(p,q)$. 

These considerations suggest that
a noncommutative structure of spacetime could be introduced starting 
with the classical 
commutative manifold equipped with a symplectic form $\Omega$. 
A path integral formalism similar to Eq.(\ref{3}) may then provide
a way of evaluating a class of linear functionals over polynomials of
the deformed noncommutative position and time operators $X^\mu$.
Of course this will restrict
our analysis to the case of even dimensional spacetime manifolds 
$M^{2n}$, in order that a 
nondegenerate, closed two-form $\Omega$ exists.
This construction requires the introduction of a parameter with dimension of 
length, $\lambda_P$, which plays a role analogous of $\hbar$ in ordinary quantum
mechanics. In particular in the limit $\lambda_P \rightarrow 0$ all 
functionals we will consider reduce to evaluation maps of the commutative
algebra of smooth functions over $M^{2n}$. For finite $\lambda_P$ instead, 
position and time operators $X^{\mu}$ satisfy non trivial 
commutation relations of the form of Eq.(\ref{1}) with $Q^{\mu \nu}$ related
in a simple way to $\Omega$.

\section{Path Integral over spacetimes}

We start considering the pair $(R^{2n}, \Omega)$, with $\Omega$ 
a symplectic form and the following {\it generating 
functional}  
\begin{equation}
Z(x_0,J) = N \int D[\gamma]~ \exp \left[ i \lambda_P^{-2} 
\left( \Sigma [x_0,\gamma]  +  (x,J) \right) \right]
\label{5}
\end{equation}
where $N$ is a normalization constant, $J$ an arbitrary source,
the integration is carried 
over all closed curves $\gamma$
in $R^{2n}$ with base point $x_0 \in R^{2n}$ and we define
\begin{equation}
\Sigma[x_0,\gamma] = \int_{\Gamma} \Omega_{\mu \nu}(x)
dx^\mu \wedge dx^\nu
\label{6bis}
\end{equation}
\begin{equation}
(x,J) = \int_\gamma x^\mu(\tau) J_\mu(\tau) d \tau
\label{7}
\end{equation}
where we have introduced a parameterization,
$\gamma$: $\tau \mapsto x^\mu(\tau)$ and $\Gamma$ is any 
two--dimensional surface with boundary $\gamma$.
Finally the antisymmetric matrix
$\Omega_{\mu \nu}(x)$ is the representation of the symplectic 
form in the coordinate basis $x^\mu$ and, in general, explicitly
depends on $x^\mu$. Here we will assume for simplicity 
that $\Omega_{\mu \nu}(x)$ is a slowly
varying function of $x$ on the scale $\lambda_P$, and we 
consider only zero order terms in Taylor expansion around the base point $x_0$
in integrals (\ref{6bis}),  $\Omega_{\mu \nu}(x) 
\sim \Omega_{\mu \nu}(x_0)$ 
This assumption simplifies the explicit evaluation
of $Z(x_0,J)$, which is our main aim in this paper, but is not necessary.
The general case will be considered elsewhere. 
In this case $\Sigma[x_0,\gamma]$ can therefore be written as
\begin{equation}
\Sigma[x_0 ,\gamma]=
\int_\gamma x^\mu(\tau) \Omega_{\mu \nu}(x_0) {d \over d \tau} 
x^\nu(\tau) d \tau
\label{6}
\end{equation}

$Z(x_0,J)$ is invariant under curve
reparameterization, as is easy to verify, provided we redefine the
arbitrary external source $J$. In the following we will
consider $\tau \in [0,1]$. 

Equation (\ref{5}) is clearly inspired by the
expression of path integral in phase space with external sources.
However it is worth pointing out that there is 
one main difference with ordinary path integral
approach to quantum mechanics of Eq.(\ref{3}), even for $q_0=q_f$.
In this case, in fact,
the variables $q$ and $p$ are treated on different grounds, since
integration is carried over all curves with boundary conditions on
positions only, while momenta run over all possible values. In case
of Eq.(\ref{5}), instead, all curves $\gamma$ should have equal
endpoints $x_0^\mu$ {\it for all } $x^\mu$'s, $\mu=1,..,2n$.

The generating functional $Z(x_0,J)$ defines a one parameter family of 
(noncommutative) algebras ${\cal A}(\tau)$ generated by $2n$ elements 
$X^\mu(\tau)$,
implicitly defined via the introduction of a set of linear functionals 
$\rho_{x_0}$, as follows
\begin{eqnarray}
& &\rho_{x_0}(P[X^\mu(\tau_1)...X^\nu(\tau_k)]) \equiv
N Z(x_0,0)^{-1}
\int D[\gamma]~ e^{i \Sigma[x_0,\gamma]/\lambda_P^2} 
x^\mu(\tau_1)...x^\nu(\tau_k) \nonumber \\
& & = \left. (- i \lambda_P^2)^k Z(x_0,0)^{-1}
{\delta \over \delta J_\mu (\tau_1)}...
{\delta \over \delta J_\nu (\tau_k)} Z(x_0,J) \right|_{J=0}
\label{8}
\end{eqnarray}
where $P$ stands for path ordering, defined analogously to time
ordering in quantum mechanics. Once defined on all polynomials in the 
$X^\mu$, Eq.(\ref{8}) can
be applied to the entire algebra of continuous functions $F(X)$ in the weak 
topology.

To better understand Eq.(\ref{8}) we consider the {\it classical}
limit $\lambda_P \rightarrow 0$. 
The integral over curves $\gamma$ is then dominated by the 
contribution
at stationary points satisfying $\delta \Sigma[x_0,\gamma]=0$, i.e. 
$\Omega_{\mu \nu}(x_0) dx^\nu/d \tau =0$. 
Since $\Omega$ is not degenerate, the leading contribution 
satisfying $x^\mu(0)=x^\mu(1)=x_0^\mu$,
is therefore given by
the curve $x^\mu(\tau)=x_0^\mu,~\forall \tau$, so that we get
\begin{equation}
\rho_{x_0}(P[X^\mu(\tau_1)...X^\nu(\tau_k)]) 
\rightarrow \left.
x^\mu ...x^\nu \right|_{x_0} + {\cal O}(\lambda_P^2)
\label{9}
\end{equation}
Thus, the maps $\rho_{x_0}$ reduce to the evaluation maps of the commutative
algebra of smooth functions over $R^{2n}$, $\rho_{x_0}: f(x) \mapsto f(x_0)$, 
independently of $\tau$, and thus completely reconstruct $R^{2n}$ as 
a topological manifold.
For finite $\lambda_P$, instead,
all closed curves whose projection on the $x^\mu-x^\nu$ plane enclose 
surfaces with area of the 
order $\Omega_{\mu \nu}(x_0)^{-1} \lambda_P^2$ give a relevant
contribution to r.h.s. of Eq.(\ref{8}), and 
evaluation maps are smeared over cells in $R^{2n}$ 
with projection on the $x^\mu-x^\nu$ plane
of this order of magnitude.
It is worth mentioning that stationary phase method requires 
Hessian to be non degenerate at stationary points.  
As for the case of wave packets of zero mass particles, this
aspect deserves a careful analysis in the case under consideration, since 
the self--adjoint extension of the operator
$\Omega_{\mu \nu}(x_0) d / d \tau$ admits zero modes. We will deal with this 
problem by using in
the following a customary $i \epsilon $ prescription which renders the above 
operator invertible. 

We start evaluating $Z(x_0,J)$ for the case $n=1$. Use of Darboux 
theorem allows for a straightforward generalization to arbitrary $n$.
In two dimensions we write $\Omega_{\mu \nu}  (x_0)$ as $\omega(x_0) 
\epsilon_{\mu \nu}$, with  $\omega(x_0)$ a non vanishing function on $R^2$. 
Furthermore
all closed curves $x^\mu(\tau)$ with endpoints at $x_0^\mu$ can be written as 
$x^\mu(\tau)= x_0^\mu + y^\mu(\tau)$, where $y^\mu (0)=y^\mu (1)=0$. 
Substituting in (\ref{5}) we get
\begin{equation}
Z(x_0,J) = N \int D[y] \exp \left[ i \lambda_P^{-2} \left( \omega(x_0) \int_0^1 
y^\mu(\tau) \epsilon_{\mu \nu} {d \over d \tau} {y}^\nu(\tau) d \tau  
 + (x_0,J) + (y,J) \right) \right]
\label{10}
\end{equation}
The operator $\epsilon_{\mu \nu} d/d \tau$ is self-adjoint with respect to 
the scalar
product (\ref{7}) on the domain of periodic two component functions 
$y^\mu(\tau)$. Its
spectrum is given by $\mu_m= 2 m \pi$, $m \in Z$. To each eigenvalue 
correspond the two 
eigenfunctions, normalized to $\lambda_P$
\begin{equation}
y^{(+)}_m(\tau)= \lambda_P \left( \begin{array}{c} cos(\mu_m \tau) \\ 
sin(\mu_m \tau) \end{array} \right)
~~~,
y^{(-)}_m(\tau)= \lambda_P \left( \begin{array}{c} sin(\mu_m \tau) \\ 
-cos(\mu_m \tau) \end{array} \right)
\label{11}
\end{equation}
Expanding the functions $y^\mu(\tau)$ in the basis (\ref{11}),  
$Z(x_0,J)$ becomes an
usual pseudo-gaussian integral over the Fourier components $c^{(\pm)}_m= 
\lambda_P^{-1}(y^{\pm}_m,y)$. 
As mentioned before, we will introduce a $i \epsilon$ prescription as a 
regularization of the
singularity due to the presence of a zero mode, namely shifting in the 
complex plane
$\mu_m \rightarrow \mu_m + i \epsilon$. In this way we get,
integrating over $c^{(\pm)}_m$
\begin{eqnarray}
 & Z(x_0,J) & = N' exp \left[ i \lambda_P^{-2} x_0^\mu \int_0^1 J_\mu(\tau) 
d \tau 
-(4 \epsilon \lambda_P^2)^{-1} \delta^{\mu \nu} \int_0^1 J_\mu(\tau) 
d \tau 
\int_0^1 J_\nu(\tau') d \tau' \right. \nonumber \\
& & \left. - i \sum_{m \neq 0} (4 \mu_m \omega(x_0) + 
i \epsilon)^{-1}
\left((J^{(+)}_m)^2+(J^{(-)}_m)^2 \right) \right]
\label{12}
\end{eqnarray}
where $J^{(\pm)}_m= \lambda_P^{-1}(x^{(\pm)}_m,J)$ are the Fourier components 
of $J_\mu(\tau)$ and   
notice that the zero mode does not contribute to the sum, so we can 
safely neglect
the $i \epsilon$ term in this case. Using (\ref{11}) and performing the 
sum over $m$ we 
finally obtain
\begin{eqnarray}
 & Z(x_0,J) & = N' exp \left[ i \lambda_P^{-2} x_0^\mu \int_0^1 J_\mu(\tau) 
d \tau 
-(4 \epsilon \lambda_P^2)^{-1} \delta^{\mu \nu} \int_0^1 J_\mu(\tau) 
d \tau 
\int_0^1 J_\nu(\tau') d \tau' \right. \nonumber \\
& & \left. + i \lambda_P^{-2} \int_0^1 d \tau \int_0^1 d \tau' 
\Omega^{-1 \mu \nu} (x_0) J_\mu(\tau) J_\nu(\tau') 
\Delta(\tau-\tau') \right]
\label{13}
\end{eqnarray}
where the kernel $\Delta(t)$ is given by
\begin{equation}
8 ~ \Delta(t)  = \left\{ \begin{array}{c} 2 t -1 
~~~t \in ~ ]0,1[ \\
~~2 t +1 ~~~t \in~ ]-1,0[ \end{array}\right.
\label{14}
\end{equation}
which is discontinuous at $t=0$.
We can now proceed to explicitly evaluate the action of the maps $\rho_{x_0}$ 
on path
ordered polynomials of $X^\mu$. Up to second order polynomials we get
\begin{equation}
\rho_{x_0}(X^\mu(\tau))= x_0^\mu~~~,\rho_{x_0}(P[X^\mu(\tau) X^\nu(\tau')])= 
x_0^\mu x_0^\nu + {\lambda_P^2 \over 2 \epsilon} \delta^{\mu \nu} 
- 2 i \lambda_P^2 \Omega^{-1 \mu \nu} (x_0) \Delta(\tau-\tau') 
\label{15}
\end{equation}
Actually, since $Z(x_0,J)$ only contains linear and quadratic
terms in $J$, all other higher order polynomials can be evaluated, up to 
combinatorial
factors, in terms of these expressions.

We see from (\ref{15}) that $\rho_{x_0}$ simply associates to $X^\mu(\tau)$ 
the value 
$x_0^\mu$, independently of $\tau$. In fact the only dependence on the $\tau$ 
variable
is contained in the function $\Delta(\tau - \tau')$, appearing for 
path ordered products of $X^1(\tau)
X^2(\tau')$. As long as arbitrary polynomials $X^\mu(\tau_1)...X^\mu(\tau_k)$ 
in only one of the $X^\mu$ operators are considered, 
the dependence on the parameters $\tau_i$ drops out. We also notice that 
we get for the {\it uncertainty} functional
\begin{equation}
\rho_{x_0} (X^\mu(\tau) X^\mu(\tau')) -
 \rho_{x_0}(X^\mu(\tau)) \rho_{x_0}(X^\mu(\tau')) = 
{\lambda_P^2 \over 2 \epsilon}~~~\forall~ \tau, \tau'
\label{16}
\end{equation}
These results have a customary interpretation by looking at 
$\rho_{x_0}$ as corresponding
to states $|x_0>$ in a Hilbert space with respect to which $X^\mu(\tau)$ 
have mean value
$x_0^\mu$ and  squared uncertainty as in (\ref{16}). 
For the equal $\tau$ commutator we find
\begin{eqnarray}
 \rho_{x_0}([X^\mu,X^\nu](\tau)) & = & \lim_{\delta\rightarrow 0^+}
\left( \rho_{x_0}(X^\mu(\tau+ \delta) X^\nu(\tau)) \right. \nonumber \\
 - \left. \rho_{x_0}(X^\nu(\tau+ \delta) X^\mu(\tau')) \right) & = & {i \over 
2} \lambda_P^2
\Omega^{-1 \mu \nu} (x_0)  ~~~,\forall \tau
\label{16bis}
\end{eqnarray}
As expected the commutator is proportional to the inverse symplectic form 
$\Omega^{-1}$ and using
notation of Eq.(\ref{1}), we see that $X^{\mu}(\tau)$ generates a 
noncommutative algebra with, for any
$\tau$, $\rho_{x_0}(Q^{\mu \nu}(\tau))= \Omega^{-1 \mu \nu}(x_0)/2$.

The uncertainties (\ref{16}) diverge if one let the regularizing 
parameter $\epsilon$ to go to zero. This means that in this limit
the states $|x_0>$ are no more in the domain of $X^{\mu}$. However
the $1/\epsilon$ singularity can be removed if one define a
{\it renormalized} set of linear functional 
$\tilde{\rho}_{x_0}$ which satisfy the first of Eqs.(\ref{15}),
Eq.(\ref{16bis}), and have minimal squared uncertainties compatible 
with Eq.(\ref{16bis}), and equals for both $X^1(\tau)$ and 
$X^2(\tau)$, namely $\lambda_P^2 (4 \omega(x_0))^{-1}$. Equation
(\ref{15}) then becomes
\begin{equation}
\tilde{\rho}_{x_0}(X^\mu(\tau))= x_0^\mu~~~,\tilde{\rho}_{x_0}(P[X^\mu(\tau) 
X^\nu(\tau')])= 
x_0^\mu x_0^\nu + {\lambda_P^2 \over 4 \omega(x_0)} \delta^{\mu \nu} 
- 2 i \lambda_P^2 \Omega^{-1 \mu \nu} (x_0) \Delta(\tau-\tau') 
\label{18}
\end{equation}
Accordingly, the corresponding generating functional 
$\tilde{Z}(x_0,J)$ is obtained 
from Eq.(\ref{13}) by substituting $\epsilon$ with $2 \omega(x_0)$.
The states corresponding to this new set of linear functionals would
be the analogous of coherent states for ordinary quantum mechanics.
They would also correspond to the states of maximal localization 
introduced in Ref.\cite{km}. Notice also that $\tilde{\rho}_{x_0}$
give finite results when applied to arbitrary higher order 
polynomials
in $X^\mu(\tau)$, which, as already stressed, can all be expressed in
terms of (\ref{18}).

The above results can be easily generalized to the $2n$ dimensional 
case. We start again with Eqs.(\ref{5})-(\ref{7}), written now on 
$R^{2n}$. As before we will consider in Eq.(\ref{6bis}) only zero terms in 
Taylor expansion of $\Omega(x)$.
By Darboux
theorem it is always possible reduce $\Omega(x_0)$, for any $x_0$, 
to the canonical form 
\begin{equation}
\omega= \left(\begin{array} {cc} ~0 ~~~~1_n \\ - 1_n ~~0 \end{array} \right)
\label{19}
\end{equation}
with a suitable coordinate transformation $z^\mu = (A^{-1})^\mu_\nu(x_0) 
y^\nu$.
In these new coordinates 
and within the approximation $\Omega(x) \sim \Omega(x_0)$,
$\Sigma[x_0,\gamma]$ is simply the sum of
$n$ bidimensional integrals like the first term in Eq.(\ref{10})
and $Z(x_0,J)$ factorizes into the product of $n$ bidimensional
generating functionals depending on the sources 
$J'_\mu = (A^T)^\nu_\mu(x_0) J_\nu$,
with $A^T$ the transpose of $A$.  
Carrying out the pseudogaussian integrals and transforming back the
result in terms of $J_\mu$ we therefore obtain
\begin{eqnarray}
 & Z(x_0,J) & = N' exp \left[ i \lambda_P^{-2} x_0^\mu \int_0^1 J_\mu(\tau) 
d \tau 
-(4 \epsilon \lambda_P^2)^{-1} (A A^T)^{\mu \nu}(x_0) \int_0^1 
J_\mu(\tau)  d \tau \cdot
\right. \nonumber \\
& & \left. \cdot \int_0^1 J_\nu(\tau') d \tau' 
+ i \lambda_P^{-2} \int_0^1 d \tau \int_0^1 d \tau' 
\Omega^{-1 \mu \nu} (x_0) J_\mu(\tau) J_\nu(\tau') 
\Delta(\tau-\tau') \right] \label{20}
\end{eqnarray}
which gives
\begin{equation}
\rho_{x_0}(X^\mu(\tau))=x_0^\mu
\end{equation}
\begin{equation}
\rho_{x_0}(P \left[ X^\mu(\tau) X^\nu(\tau') \right]) =
x_0^\mu x_0^\nu + \frac{\lambda_P^2}{2 \epsilon} (A A^T)^{\mu \nu} (x_0)
- 2 i \lambda_P^2 \Omega^{-1 \mu \nu} (x_0) \Delta(\tau - \tau')
\end{equation}
and for the equal $\tau$ commutator again Eq.(\ref{16bis}).
 
It is worth noticing that a constant $\Omega$ in four dimensions
corresponds to the case considered in Ref. \cite{dfr}, 
where the following algebra has been studied
\begin{eqnarray}
[X^\mu,X^\nu] & = & i \lambda_P^2 Q^{\mu \nu},~~~
[X^\mu,Q^{\mu \nu}]=0, \nonumber \\
Q_{\mu \nu} Q^{\mu \nu} & = & 0,~~~\left( {1 \over 8} 
\epsilon_{\mu \nu \rho \sigma} Q^{\mu \nu} Q^{\rho \sigma} \right)^2=I
\end{eqnarray}
Actually the
condition $\rho_{x_0}([X^\mu,Q^{\nu \rho}](\tau))=0$, $\forall~ \tau$ 
can be easily verified using Eq.(\ref{20}), while the constraint 
$\rho_{x_0}(\epsilon_{\mu \nu \rho \sigma} 
Q^{\mu \nu} Q^{\rho \sigma}(\tau))= \pm 8$
is equivalent, up to a suitable normalization, to require $\Omega$ to be non 
degenerate. The procedure of describing a noncommutative geometry with a path 
integral approach outlined in this paper cover however more general cases,
since the commutator may depend on $x_0$, as clear from Eqs.(\ref{16bis})
and (\ref{20}). Moreover, if one
does not assume that $\Omega(x)$ is a slowly varying function 
of $x$ and that for each $x_0$ one can approximate 
$\Omega(x) \sim \Omega(x_0)$ in the phase $\Sigma[x_0,\gamma]$,
one may consider, at least in principle, a completely
general case, 
though the corresponding generating functional could be
difficult to handle. 
Nevertheless, even if $Z(x_0,J)$ is not exactly solvable
for more involved choices of $\Omega$, applications of perturbation techniques,
as in ordinary path integral in quantum mechanics, may provide some information
on the underlying noncommutative spacetime.

\section{Path Integral over spacetimes with non-trivial topologies}

In the previous section we have considered the path integral construction of
noncommutative generalizations of spacetimes for 
the case of $R^{2n}$ as the underlying classical manifold. When the method 
outlined there is applied to manifolds $M^{2n}$
with non trivial topologies, in 
particular when the first and/or second homology groups are non trivial, there
are however two aspects which one should take into account.
First of all, if $H^1(M^{2n}) \neq \{ 0 \}$, there are 
closed curves contributing
to the sum over paths which are not boundaries of any surface, so the 
surface integral of the assigned symplectic form is not defined for these paths.
In these cases the introduction of a one-form $A$, such that locally 
$\Omega = dA$ is a better starting point.
Moreover, if $H^2(M^{2n}) \neq \{ 0 \} $, 
the integral  $\Sigma[x_0,\gamma]$ for any
$\gamma$ is in general a multivalued function, since the integral of $\Omega$ 
over two surfaces $\Gamma$
and $\Gamma'$, both with boundary $\gamma$, may be different if 
$\Gamma-\Gamma'$ is a non trivial 2-cycle (of course this problem is absent
if $\Omega$ is exact). The ambiguity in the choice of the
surface, however, can be removed by requiring {\it quantization} conditions
on the integral of $\Omega$ over a set of generators of $H^2(M^{2n})$
such that $exp(i \Sigma[x_0,\gamma] \lambda_P^{-2}$) is
single valued, analogous
to the one introduced to quantize the motion of a charged 
particle in the field of a magnetic charge.
We will illustrate these points in two bidimensional simple examples, the 
sphere $S^2$ and the torus $S^1 \times S^1$, and discuss the generalization
to general cases.

Let us consider the pair $(S^2,\Omega)$. Any closed curve $\gamma$ 
will divide $S^2$ into two surfaces, and therefore the value of
$\Sigma[x_0,\gamma]$ of Eq.(\ref{6bis}) will be defined only up to
the integral of $\Omega$ over $S^2$. The measure in the path integral will
be nevertheless single valued if the latter is a multiple of $2 \pi$, in unit 
$\lambda_P^2$
\begin{equation}
\int_{S^2} \Omega = 2 n \pi \lambda_P^2
\label{quant}
\end{equation}
A path integral over $S^2$ was already considered in
\cite{nr}, as a way to quantize a classical particle with spin. 

Incidentally, we observe that
condition (\ref{quant}) could have an intriguing relationship with the idea that
black hole horizon area is quantized and its spectrum is uniformly spaced
\cite{bek}. In our approach this quantization, in unit of $\lambda_P^2$,
would be intimately related to the fact that positions and time are noncommuting
operators and their commutator is of the form reported in Eq.(\ref{1}).
Actually black hole physics is one of the best scenarios where the
noncommutative nature of spacetime at small scales, otherwise unobservable,
could be felt well above the Planck length. 

The ambiguity in the definition of the measure in path integral and the
necessity of a {\it quantization} condition (\ref{quant}) is due to the
fact that the second homology group of $S^2$ is non trivial. 
A similar ambiguity in defining the {\it action} $\Sigma[x_0,\gamma]$
corresponding to a path
$\gamma$ on a manifold $M^{2n}$, and therefore the generating functional 
(\ref{5}), will arise whenever $H^2(M^{2n})\neq \{ 0 \}$.
In particular, if its 2-Betti number is equal to $b_2(M^{2n})=k$, and denoting
with $\Gamma_i$, $i=1,...,k$, a set of generators of $H^2(M^{2n})$, the measure
$exp( i \Sigma[x_0,\gamma] \lambda_P^{-2} )$ will be single valued if $\Omega$
satisfies the following conditions
\begin{equation}
\int_{\Gamma_i} \Omega = 2 n_i \pi \lambda_P^2~~~, i=1,...,k
\end{equation}
which of course do not depend on the choice for the elements of the basis
$\Gamma_i$.

As a second example of a non trivial topological manifold, we consider the
bidimensional torus $T^2=S^1 \times S^1$.  In this case there are closed curves
which are not boundaries of any surface and therefore the construction 
outlined in the previous section cannot be applied.  
Nevertheless it is still possible to construct a path integral over $T^2$ 
and the generating functional $Z(x_0,J)$ starting with a one--form $A$
such that (locally) $\Omega=d A$. 
Let us denote with $\gamma_i(x_0)$, $i=1,2$ two generators of the $T^2$ 
1-homology group passing through the point $x_0$ and define
\begin{equation}
I_i(x_0)= \lambda_P^{-2} \int_{\gamma_i(x_0)} A 
\label{cyclei}
\end{equation}
The path integral measure
we will introduce soon will be single valued only
if the values of $I_i(x_0)$ only differ for a factor $2 n \pi$
when the curve $\gamma_i$ is continuously moved around the manifold and go back
to the original path after a complete cycle. For example, in the angular 
coordinates $\theta$ and $\phi$ on $T^2$, taking $A=r_1 r_2 
( \theta d \phi - \phi
d \theta)$,with $r_1,r_2$ the two radii,
and choosing $\gamma_1(\theta_0,\phi_0) : \{ \theta= \theta_0 \}$ and
$\gamma_2(\theta_0,\phi_0) : \{ \phi = \phi_0 \}$, the values of $I_1(x_0)$
for the identified points $\theta_0$ and $\theta_0+2 \pi$,
and analogously for $I_2(x_0)$, will be the same up to a factor $2 n \pi$ 
provided that
\begin{equation}
\int_{T^2} d A = 4 \pi^2 r_1 r_2 = 2 n \pi \lambda_P^2
\end{equation}
This condition also ensures, as we have seen for the $S^2$ case, that for
all curves which are instead boundaries of surfaces the corresponding 
contribution to the generating functional is single valued.
Defining 
\begin{equation}
\Sigma [x_0,\gamma] = \int_\gamma A
\end{equation}
the contribution to $Z(x_0,J)$ of a generic path $\gamma$ on $T^2$ can be
written as
\begin{equation}
\exp \left[ i \lambda_P^{-2} \Sigma [x_0,\gamma] \right] = 
\exp \left[ i n_1 I_1(x_0) + i n_2 I_2(x_0) + i 
\lambda_P^{-2} \int_{\gamma'} A \right]
\end{equation}
for some integers $n_1$ and $n_2$ and with $\gamma'$ a boundary of some
bidimensional surface, so the generating functional takes the form
\begin{equation}
Z(x_0,J) = N \sum_{n_1,n_2} \int D[\gamma'] 
\exp \left( i n_1 I_2(x_0) + i n_2 I_2(x_0) \right)
\exp \left( i \lambda_P^{-2} \int_\Gamma dA \right)
\exp \left(  i \lambda_P^{-2} (x,J)  \right)
\end{equation}
where the integration is over all 1--boundaries $\gamma'$ 
with base point $x_0$ and $\partial \Gamma = \gamma'$.

This expression can be now easily generalized to an 
arbitrary manifold $M^{2n}$ with Betti numbers $b_1=p$, $b_2=q$.
Consider the pair $(M^{2n},A)$, a set of basis $\gamma_i(x_0)$ and 
a base $\Gamma_j$,
for $H^1(M^{2n})$ and $H^2(M^{2n})$ respectively. 
The fact that the path integral measure should be single valued enforces
$q$ integral conditions on $\Omega = d A$
\begin{equation}
\int_{\Gamma_j} \Omega = 2 m_j \pi \lambda_P^2,~~~j=1,...,q
\end{equation}
and defining, as in (\ref{cyclei})
\begin{equation}
I_i(x_0) = \lambda_P^{-2} \int_{\gamma_i(x_0)} A,~~~ i=1,...,p 
\end{equation}
we may introduce the generating functional as follows
\begin{equation}
Z(x_0,J) = N \prod_{i=1}^p \sum_{n_i} \int D[\gamma'] 
\exp \left( i n_i I_i(x_0) \right)
\exp \left( i \lambda_P^{-2} \int_\Gamma \Omega \right)
\exp \left( i \lambda_P^{-2} (x,J) \right)
\label{general}
\end{equation}
where $\partial \Gamma = \gamma'$. 
Equation (\ref{general}) represents the generalization of Eq.(\ref{5}) to
a generic topological manifold $M^{2n}$.

\section{Conclusions and outlook}

In this paper we have described a path integral approach to noncommutative
geometries. In particular, using the analogy with ordinary quantum
mechanics we have described a way of expressing a 
set of linear
functionals over noncommutative generalizations, in even dimensions, 
of the algebra generated by position and time
operators in terms of ordinary functions over the classical manifold,
which represents its commutative {\it classical} geometry. In particular all 
these linear functionals reduce 
in the limit $\lambda_P=0$ to evaluation maps 
over the commutative algebra of smooth functions over the manifold.

There are several points related to this approach 
which deserve further studies. 

We have considered in more details the simple case of a classical 
starting manifold with trivial first and second homology groups. 
It would be interesting to perform  
a deeper study of the properties of the generating functional over 
manifolds with more involved topological properties, which has been discussed 
in section 3, and of the corresponding deformed noncommutative geometries.

Another important issue is that this approach may give 
the possibility of introducing a differential calculus over
noncommutative generalizations of the algebra of functions 
over classical spacetimes 
using all customary results for a commutative 
geometry. In particular,
quite relevant is to look for a consistent definition of a metric structure,
connections and curvature. Of course, when $\lambda_P$ is switched on, one 
expects that
dramatic changes affect all these structures. Nevertheless, to have a 
simple way of 
relating the algebra generated by the $X^\mu$'s to its commutative 
counterpart, 
could be a powerful way to look for their generalization.

The role of the parameter $\tau$ should be better understood. While it has 
a clear
meaning as the parameter of closed curves over $M^{2n}$, its role for the 
noncommutative
algebras ${\cal A}(\tau)$ is not transparent. Actually all dependence on 
$\tau$ is 
contained in the kernel $\Delta(\tau-\tau')$.
One possibility is of course that each of these algebras, for fixed $\tau$,
is already sufficient to describe a noncommutative spacetime, so that the role
of the parameter $\tau$ in this case would be only auxiliary. On the other hand 
the analogy with quantum phase space we started with, suggests that it could 
also have an intrinsic non trivial meaning.

\section{Acknowledgments}
 
I am pleased to thank F. Lizzi, G. Miele and G. Sparano for 
discussions and comments.

\end{document}